# From Search Engines to Search Services: An End-User Driven Approach

Gabriela Bosetti[1], Sergio Firmenich[1,2], Alejandro Fernandez[1], Marco Winckler[3], Gustavo Rossi[1,2]

[1]LIFIA, CIC, Facultad de Informática, Universidad Nacional de La Plata
`{gabriela.bosetti, sergio.firmenich, alejandro.fernandez, gustavo}@lifia.info.unlp.edu.ar`

[2] CONICET, Argentina

[3] ICS-IRIT, University of Toulouse 3, France
`winckler@irit.fr`

**Abstract.** The World Wide Web is a vast and continuously changing source of information where searching is a frequent, and sometimes critical, user task. Searching is not always the user's primary goal but an ancillary task that is performed to find complementary information allowing to complete another task. In this paper, we explore primary and/or ancillary search tasks and propose an approach for simplifying the user interaction during search tasks. Rather than focusing on dedicated search engines, our approach allows the user to abstract search engines already provided by Web applications into pervasive search services that will be available for performing searches from any other Web site. We also propose to allow users to manage the way in which searching results are displayed and the interaction with them. In order to illustrate the feasibility of this approach, we have built a support tool based on a plug-in architecture that allows users to integrate new search services (created by themselves by means of visual tools) and execute them in the context of both kinds of searches. A case study illustrates the use of these tools. We also present the results of two evaluations that demonstrate the feasibility of the approach and the benefits in its use.

## 1 Introduction

Searching is one of the main important tasks that users perform when using Web browsers. In terms of user aims, searches in the Web may occur as a main task (primary search) or as a secondary one (ancillary search) [2]. According to this work [2], primary search corresponds to the user's primary need for information, usually involving a single cycle question-answer (e.g. looking for something in Google or Amazon). In opposition, ancillary searches are aimed at providing details (under demand) about the current information been displayed in a Web site the user is accessing; this task can be performed in different ways, e.g. by using an external Web page's search engine or the contextual search menus that Web browsers provide. In order to delve deeper into the search process, we show in Figure 1 two task models: one for primary searches and other for the ancillary ones. Figure 1.a depicts how primary searches are independent of each other (e.g. finding the Web site of ICWE2016 or the WWW2016, in both cases, they are independent primary searches). In opposition, ancillary searches may become a recursive task, but in the background the main focus is still the original Web site. Figure 1.b depicts an example of this kind of search. Here, a researcher is looking for authors and articles in the ICWE2016 Web site. If this user requires further information about one of the authors, he can look for him at DBLP. There, the user may be interested in some specific resulting article, which triggers another ancillary search in Google Scholar to find the file or see the amount of citations. After that, the user may require to repeat a similar cycle of ancillary information seeking, i.e. search for another author/article listed in ICWE2106's Web site. Note that this last scenario may imply a detriment of user experience because it requires moving information and its context from one search Web site to another one. This get worse given the proliferation of Web applications managing millions of information items from diverse domains (cinema, tourism, wikis, Web searches, etc.) that could be queried by users.

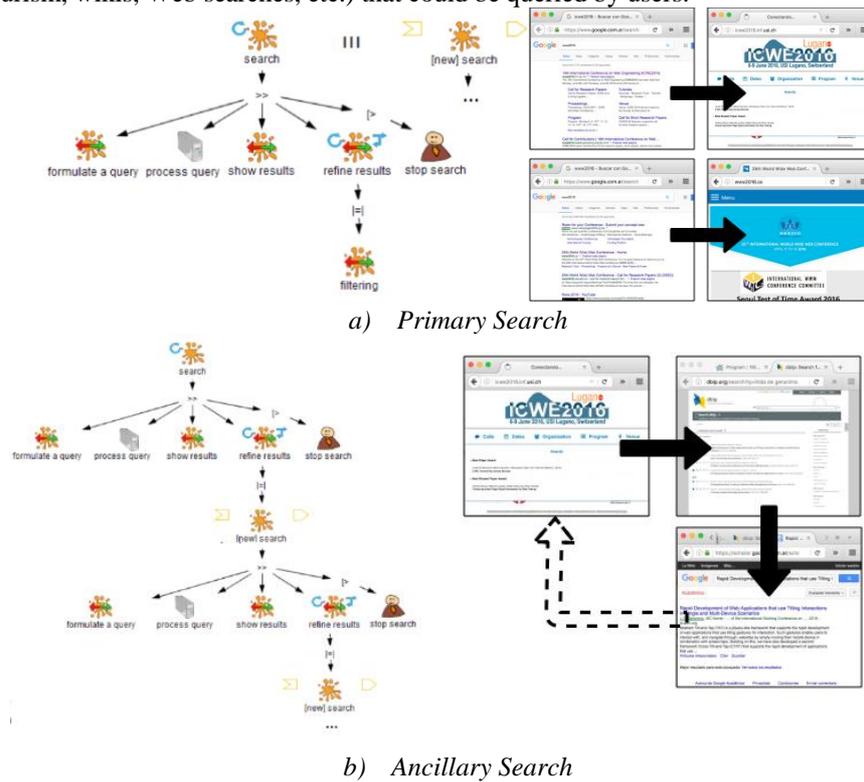

*a) Primary Search*

*b) Ancillary Search*

**Fig. 1**: Primary and ancillary search task models and scenarios

It must be clear that the information required by users in both kind of searches may be really different [13]. In ancillary searches, users need to move among Web sites just for obtaining complementary information that is frequently domain-specific (e.g. computer science literature), and consequently they would choose to use a specific Web site to look for such kind of information (e.g. DBLP or GoogleScholar). A second aspect in this kind of search is how and where the interaction with search results occurs. Usually, the interaction with search results is in the external Web page where the search was actually performed and not in the search-execution context (e.g. the Web page from where the user was motivated to perform the ancillary search).

Here, it is important to differentiate between generic Web searchers (based on Web content scrapers) and domain-specific ones (based on domain objects). For performing primary search, it is common that users prefer generic Web searchers (such as Google), but for ancillary ones, it is probable that the user prefers to use an information repository that actually store information for a specific domain. When the user knows that they need domain-specific information, generic Web searchers are not much useful.

In this situation, simple and frequent search tasks are hardly supported by Web browsers, which are integrated by default with well-known searchers. For instance, Safari allows users to specify the searcher to use among Google, Yahoo, Bing or DuckDuckGo. A more flexible strategy is followed by Mozilla Firefox. In this case, the search widget presented at the top bar may be extended in two ways. First, developers may create specific kind of Firefox extensions that end-users may install and use. Second, the end-user may choose to add a particular search when Firefox detects that the current Web page provides one. However, this last feature works only for very well-known Web sites (such as IMDB), but not for every Web site providing search functionality. This lack of integration is understandable given that, in the current state of the Web with a huge amount of Web applications and end-users, browser vendors are not able to support every users' expectations in relation to integrated Web searchers.

In this paper, we propose and end-user driven approach in which users are empowered with the possibility of abstracting those search engines provided by their preferred Web applications into Web Search Services. These services are then deployed to allow users to trigger ancillary searches and integrating the results in the current Web site. Results are not just presented in-situ, but visualized in different ways and it is also possible to interact with them from such context (e.g. ordering, filtering).

An overview of this idea is shown in Figure 2. In this case, we show how our approach would better support the scenario presented in Figure 1.b. Instead of requiring to change the Web site in use, we propose to obtain search results transparently for the user and show them in the same context of the current Web site, which will make easier and faster to repeat the same search process with other authors and enable the possibility of comparing their results without moving among tabs or windows.

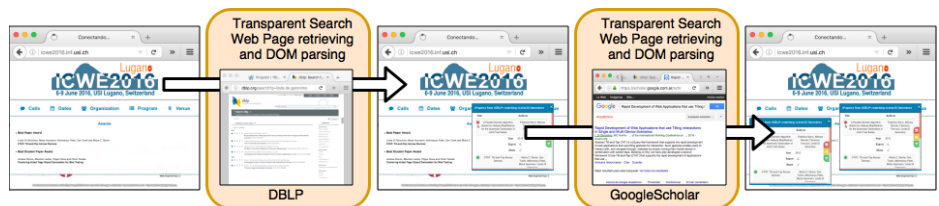

**Fig. 2**: Search Approach scenario

In order to make this possible for all users, we provide an end-user development tool for the creation of these Search Services that will be integrated automatically for further uses. These Search Services are specified in terms of a model designed to this aim, allowing also to share the produced service specifications among users.

The rest of the paper is organized as follows. First, we present and discuss the related works. Then we present our approach, the underlying architecture and tools. In Section 4 we explain deeply our support tool through a case study. An evaluation showing the feasibility of abstraction of Search Services is presented in Section 5, where we also present the results of a quantitative assessment. Finally, we discuss our approach, highlight some final remarks and present the future works.

## 2 Related works

Our approach has several facets related to different topics in Web search. First, in this section, we go through the discussion about information seeking in the context of user tasks. We also discuss about the importance of the semantic structure of results. The reader should consider an underlying crosscutting aspect in this section, that is the support of secondary or ancillary searches, which is also particularly discussed at the end.

Web search has been a target of research from the beginnings of the Web. Several works have contributed in different aspects: scraping, semantic search, collaborative search, etc. In spite of that, a deeper understanding of ancillary searches was recently made [2], something really relevant at the moment of taking decisions.

As early as 1999, Ford and colleagues [4] studied the implications evidence based practice (EVP) for information retrieval. Evidence-based information retrieval implies, for example, techniques that allow the user to indicate the extent to which retrieved sources should have been critically evaluated, or to indicate their own evaluation criteria. Although EVP was originated in the medical sciences, it spread to other disciplines such as the social sciences and education. Sadly, these disciplines still rely on general-purpose information retrieval systems such as citation databases and search engines. The functionality we describe in this work improves the situation in three ways. Firstly, it adapts various information sources to a common (objectified) interface that allows rich and more complex searches over various sources. Secondly, it allows the user creating the Search Service to define quality filters (in terms of information available) to be applied before any results are returned. Thirdly, the metadata of each Search Service allows rich classification of information sources/repositories. Future work will explore strategies to automatically bind specific properties to open vocabularies and ontologies such as DBPedia (e.g., via DBPedia Spotlight[1]).

Research on information seeking behaviour has focused mostly on how individuals seek information. However, in many contexts, information seeking involves collaboration [5]. The Search Service definition tool described here stores service definitions in a local storage. Users can export and import definitions thus supporting simple sharing via e-mail and instant messaging. MacLean and colleagues found this simple form of collaborative tailoring via customization files and email sharing an effective mechanism to foster a culture of end-user tailoring [6].

---

[1] https://github.com/dbpedia-spotlight/dbpedia-spotlight - last accessed on Dec/01/2016

Web search engines return a result page with a list of Web documents (URIs) matching the search criteria. Results are usually presented as a list of page titles and one or two sentences taken from the content. Recent advances in Web search interfaces provide, for a predefined set of element types such as movies, recipes, and addresses, rich snippets that help the user recognize relevant features of each result element (e.g., movies playing soon or user opinions for a given movie). In some cases, these rich snippets include the data the user is looking for. They are possible because Web site creators include structured data in Web pages that computers can recognize and interpret, and that can be used to create applications. Viewing the Web as a repository of structured, interconnected data that can be queried is the ultimate goal of the Semantic Web [7]. However, end users do not always have the means to exploit or add information to the Semantic Web. The tools presented here let end-users without training on Web development to extract and use structured data from any Web site.

There are models that describe how people deal with information and their information needs [8], and consequently provide context to the tools we propose. They found out that eight characteristics (features) adequately describe information behaviour in various disciplines [9]. Such characteristics are: starting (means used to begin seeking information), chaining (following connections to related material), browsing (semi-structure browsing), differentiating (using known differences to filter information), monitoring (keeping up-to-date), extracting (selectively identifying relevant material in an information source), verifying (checking for accuracy), and ending (final searches and tidying up). The definition of Search Services allows the extraction of relevant features in Web content, putting the user in control of what are the relevant properties of information objects, to latter use them for differentiation. Moreover, when users select which information repositories they will define search services for, they conduct an initial quality filter (verification by provenance). Being able to mouse-select elements in the primary window to launch in-context searches is a means to support search start and information chaining activities. When the results are displayed in overlay mode, these connections become explicit and persistent, as it is pointed in [2]. However, although [2] motivates the need of an inside-in search approach, the proposed system is not oriented to end-users, but that it retrieves results from a single broker that is integrated by predefined databases. Although we share the philosophy of inside-in searching, our work is on the technical characteristics of the system required to allow end-users to define their own search services. With this in mind, our approach may be integrated with any existing search engines, without depending on a specific broker. In this context, beyond how to abstract these search engines, another necessary difference with [2] is that are users how must define the semantic and structure of results.

## 3   End-User Driven Search Services

In this section, we present our approach for the integration between Web browsers and Web searches. First, we introduce a deeper explanation of the steps composing a search task, and then, we focus on the steps where our approach intervenes. Following, we present the underlying architecture, the model for Search Services specification and the role of our tools, which are oriented to be used by end-users.

### 3.1 Contextualizing our approach

In order to better understand the contribution of our approach, it is necessary to delve deeper into the full process of searching from the end-users' point of view. Without taking into account if it is concerning a primary or an ancillary search, this process involves the following steps: 1) define a query, 2) select a search engine, 3) entry query and trigger the search and 4) inspect and interact with the results.

Among different possible searching scenarios, we may easily appreciate that the integration support that Web applications and Web browsers provide in combination may jeopardize the user experience, since he may need to repeatedly perform extra operations (such as open a new tab, enter the URL of the target Web searcher, etc.) to obtain the desired information. Furthermore, if it was an ancillary search, the user may require refining the query or going back to the original information context.

This paper presents an approach based on a flexible architecture that allows users to customize the way they perform Web searches. Our work aims to improve the user experience related to how steps 2-4 are performed. To better support these steps, our approach is based on the following features:

1. Trigger searches from the current Web site for reducing the interaction required to perform a search in any foreign search engine.
2. Transform search results (DOM elements) into domain objects with specific semantic and structure.
3. Integrate the resulting domain instances in the current Web site for further visualization and interaction.

In order to achieve theses objectives, we propose:

1. Allowing users to encapsulate existing Web applications' search engines into pervasive Search Services. Given that not every Web application supporting searches provides an API, we propose to automatically reproduce the UI interaction required to perform a search. This implies that users must select the UI search engine components to create Search Services.
2. Integrating the new Search Services with the Web browser search mechanism for ancillary searches. Users should be able to use the created Search Services from any other existing Web site.
3. Displaying results in the context of the current Web site, in order to enable different ways of visualizations supporting primary and ancillary searches. It is done by parsing the DOM, extracting the search results from it, and creating domain object instances.

To better explain our approach, we present a simple scenario related to Figure 2[2]. There, a user is navigating through the ICWE2016's Web site, where accepted articles are listed. At some point, this user may require to see other publications of a particular author. Certainly, this secondary information requirement would be better satisfied by using domain specific Web applications (such as DBLP or Google Scholar) instead of a generic searcher (such as Google or Bing). With this in mind, and considering the use of our approach, the user would be able to trigger the search from the current Web site by highlighting the author's name and selecting the desired Search Service (in the case of the example, the results are obtained from DBLP).

---

[2] A demo for this and other scenario may be seen at: goo.gl/ljL5ey

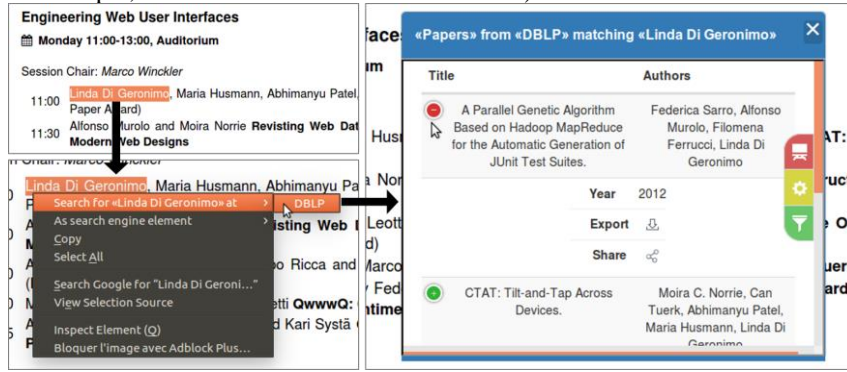

**Fig. 3**: Example of use of Search Service

The reader may note at the right that the results are listed in an overlay popup, whose content follows a table structure, which is built automatically given the domain object specification that was done when the user created the Search Service. There is also a toolbar where options for filtering and ordering results are available, if these where defined for the triggered Search Service. Furthermore, a third option allows the user to change the visualization, such in case that this table structure is not the more useful results visualization. All these aspects of our approach are presented later in this paper.

### 3.2    The approach in a nutshell

We have designed and developed an architecture for supporting the creation of Search Services and the integration tool needed for letting end-users to use these new services from any Web site they are visiting. The architecture has three layers: i) end-user support tool, ii) current search results, iii) model layer, as it shown in Figure 4.

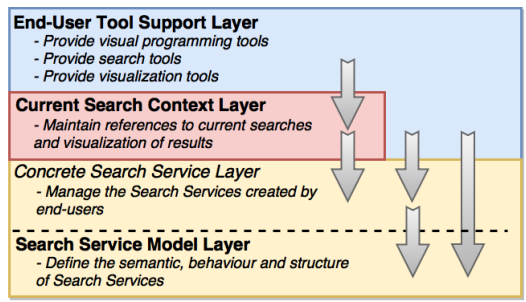

**Fig. 4**: Search Service Architecture Overview

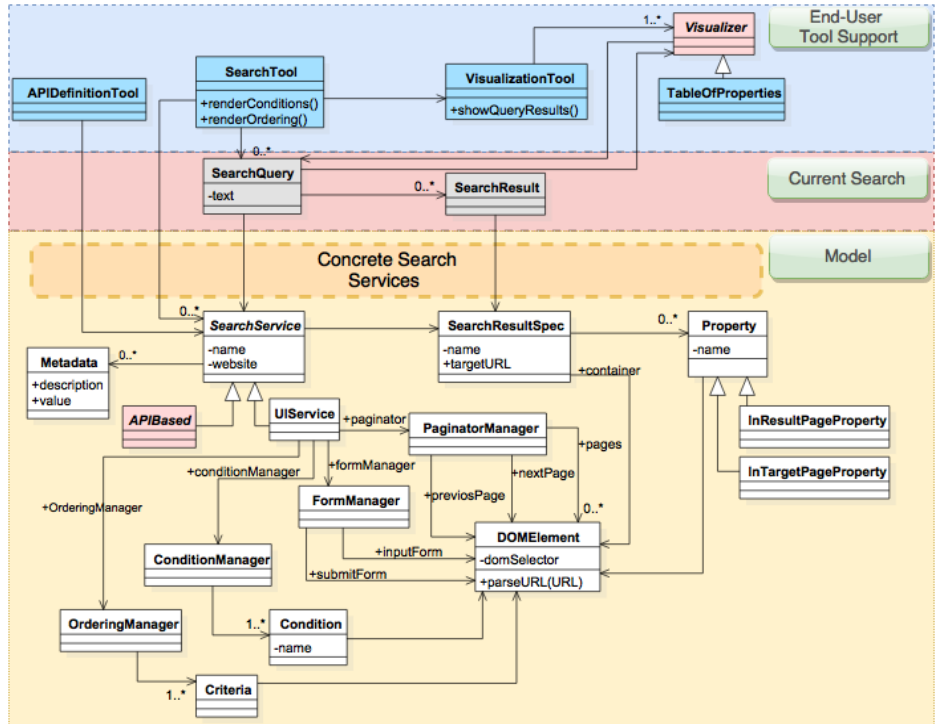

**Fig. 5**: End-User Driven Search Service Architecture

This diagram refines the relationships between layers' components:

1. *The Search Service Model*. This layer supports the creation of Search Services, which will be based in the search engines that Web applications already provide to users. Note that this is focused mainly on the creation of Search Services for those Web applications that, in spite of providing a search engine, do not provide an API. In this sense, our approach is based on the creation of a service (API alike) that is based on how users would use these search engines. Typically, a Web application search engine interface is composed by a Web form (inputs and submit button), options to filtering results and ordering them, and finally some mechanism for paginating results. We propose to abstract all these UI components (DOM elements) conforming the search engine into objects that conform a Search Service. Then, these Search Services are able to emulate the user behaviour and retrieve the corresponding results given a particular query specification. In order to provide an API-alike mechanism, the results are not just DOM elements, but also abstractions of the underlying domain objects. For instance, if the search engine being abstracted is DBLP, then the results may be wrapped in the "Paper" domain object, which could be populated with domain properties such as "title", "authors", etc. Even more, this object may have attached also properties whose values are taken from another DOM (a DOM obtained from another URL), such as a "bibtext" property; we explain this in Section 3.3. All these concepts are materialized in the bottom part from Figure 5. As we will show later, it is convenient to provide a semantic layer on top of search results, because it allows the creation of visualizers that go beyond presenting the raw results.
2. *End-User Tool Support.* These Search Services are specified by means of a set of visual tools that allow users to create them incrementally by selecting DOM elements related to the search engines of Web applications. This tool is part of the End-User Support Tool layer, where other two tools coexist. One is the *SearchTool*, which creates the corresponding menus for the existing *SearchService* objects and allow users to perform searches from the current Web site. Finally, the *VisualizationTool* takes the search results and allows users to interact with them in different ways by selecting a particular *Visualizer*.
3. *Current Search Context*. The middle layer represents the current *Query* and its results. Basically, when the user wants to perform a search task, a *SearchQuery* is created (there are basically two strategies for doing that: text selection and text input). When the search results are retrieved, these are materialized as instances of *SearchResult*; the Visualizer accesses these instances to display them to the user considering the corresponding *SearchResultSpec;* this specification describes results by means of properties whose values are also obtained from the DOM where actually the results are, or from the DOM obtained from the *targetURL*, which is basically the real domain object's Web page.

### 3.3 Search Service Architecture: Flexibility, Compliance and Extensibility

In this subsection we explain two important aspects of the architecture. Flexibility and compliance are related to how the Search Service may be mapped to different search engines provided by Web applications. Extensibility is related to how the whole toolkit may be extended with further components for searching and visualizing results.

#### 3.3.1 Flexibility and Search Engine coverage

There are some parts of the components presented in Figure 5 that deserves a bit more explanation:
- *Properties*: these are the properties defined for the *SearchResultSpec*. For instance, if the Search Service is being defined in the e-commerce domain, then properties such as *price* and *availability* could be defined. However, it must be noted that although there are some information obtainable from the DOM presenting the results, more information could be available in the actual product's site; for instance, *technical description* or *shipping costs.* Both kind of properties are extracted from different DOMs, and the behaviour for retrieving both DOMs is different. This is the reason why we separate two kind of properties. On the one hand, *InResultPageProperty* allows users to define properties whose values will be extracted from the search result's DOM. In the other hand, *InTargetPageProperty* let users to extract values for further properties. Our toolkit knows how to reach the actual result's Web page because the attribute *targetURL* is defined, mandatorily, in *SearchResultSpec*. With this in mind, we can note that when using our approach for an ancillary search, such as in the example from Figure 3, the listed properties could be obtained from different Web pages, but this will be transparent from the users' point of view.
- *Ordering and Filtering*: in order to better reproduce the power of original Web applications search engines, we have contemplated Ordering and Filtering features. Some Web sites, such as DBLP, let the users filter results according to some criteria, such as "*Journal only*", which is applied by clicking an anchor in the search page. Our Search Service model contemplates this kind of filters by means of the *ConditionManager* and *Condition* classes. All this functionality (regarding ordering and filtering), extracted from the original Web site at the time of creating the Search Service, will be available in the menu shown at the right in Figure 3. Similarly, ordering behaviour is possible to be abstracted. Most of the search engines we analysed offer ordering by clicking anchors or buttons; for instance, in Youtube, users may order results by "*Upload date*", "*Views*", etc. These ordering options are available from the same main menu in our results *Visualizer*.
- *Search Execution Strategies:* Search engines spread on the Web have different configurations concerning the involved UI components and the interaction design for executing a search: most of them have an input element to write the query but not all of them have a trigger (e.g. anchor, button) and that makes it necessary to have different strategies for carrying it out. We have implemented three alternatives, but they can be extended and integrated in the tool. We covered sites requiring to write a query into an input and handling the query execution by one of these alternatives: 1) clicking on a trigger to load the page with the results; 2) clicking a DOM element but loading the results through ajax-call; 3) listening to some input-related event (e.g. keypress, blur, change) and loading the results by ajax-call. Subclasses implementing this behaviour are *WriteAndClickToReload*, *WriteAndClickForAjaxCall* and *WriteForAjaxCall* respectively. The strategy is automatically assigned when the user creates a search service, and each strategy can know if it is applicable based on the components the user has defined and the success on retrieving new results, so this is not a concern for the end user.

*3.3.2    Extensibility*

As we show in Section 5, our Search Service Model is compliant with most of the Web sites' search engines that we analysed. However, beyond this model and the Search Service specification tool, our approach is extensible in two ways. On the one hand, the model is extensible by means of the creation of services based on the existence of application APIs. On the other hand, the end-user support tool is extensible by the creation of new Visualizers. We next explain both *APIBased* and *Visualizer* extension points.
- *APIBased Search Services*: Some Web applications offer APIs for retrieving information from their databases (Twitter, Facebook, etc.). In these cases, it is very common that APIs expressivity goes much further than what is possible to do with our UI-based Search Service approach. With this extension point, developers could create new Search Services (using application's API) to be integrated later by end-users.

- *Visualizers*: as shown in Figure 3, in our approach search results become domain objects whose properties are listed in the default Visualizer (*TableOfProperties*), which create a table where the columns represent each object' property and the rows each instance. However, further visualizers could be developed and integrated in our toolkit. For instance, a new Visualizer (*GroupByPropertyValue*) could be created in order to allow users select a property and group the already obtained results according to the value of this property. Using one more time the example from Figure 3, it would be possible to group the author's articles within *journal* or *conference* boxes. Beyond this, other visualizers could be focused on calculating information and visualizing more processed information. For instance, instead of displaying each article, the users could be interested in seeing quickly the evolution of the author production in a chart showing the amount of articles per venue.

It is important to note that these two extension points require advance JavaScript programming skills to be developed, but once created, they can be installed and used by end-users, who can configure the parameters of such new visualizers according to specific properties defined for the *SearchResultSpec* of a given *SearchService*.

## 4    Tool Support and Case Studies

In order to support our approach, we implemented our tool as a Firefox extension. It allows both the specification and execution of Search Services. The use of this tool for the first purpose is illustrated in Figures 6 and 7, and for the second one in Figure 7.

End users can define *UI-based Search Services* through our tool. A Search Service of such nature should be capable of automatically emulating a search that otherwise the user has to do manually (e.g. opening a new tab, navigating to the search engine of a Web site, typing a text, triggering a button, etc.). Once such Search Service is defined, the user can use it for performing ancillary searches by highlighting a text in any Web page he is navigating and choosing a service from which he wants to retrieve results. In this sense, the selected *UI-based Search Service* must know: in which *input control* the text should be entered, which *button* to trigger to perform the search, how to obtain more results, and how to interpret them. Filters and sorting mechanisms can also be defined, but they are not mandatory.

Consider Amaru, she is always surfing the Web and she uses to look for related books when she finds something (a topic, an author, etc.) of her interest. She is an active user of GoodReads and every time she finds some term of her interest, she copies it, opens a new tab in her browser, accesses GoodReads and performs a search with the copied text as keywords. In this setting, it would be very convenient for her to be able to carry out such searches from the same context in which she is reading the comments.

Figure 6 shows how Amaru is starting to create a *UI-based Search Service* by selecting DOM elements from the Web site of *GoodReads*, concretely from its search engine[3]. To do so, she navigates to the Web page where the search engine is and enables the «Search Service definition mode» by clicking the highlighted button in step1. In this concrete case, she should select, at least, the input (step 2) and trigger (step 4) controls, and also the one retrieving more results (step 5). The DOM elements defined as the UI-Search-Service controls are selected by right-clicking them. As this is the first time the user defines a Search Service and she has no other Services already imported in her personal account, the tool asks her to give it a name through a form opened in the sidebar, as shown in step 3. Otherwise, it will ask the user to select an existing UI-based Search Services for which it is starting to define the UI- Search-Service controls.

---

[3] https://www.goodreads.com/search?q=Borges

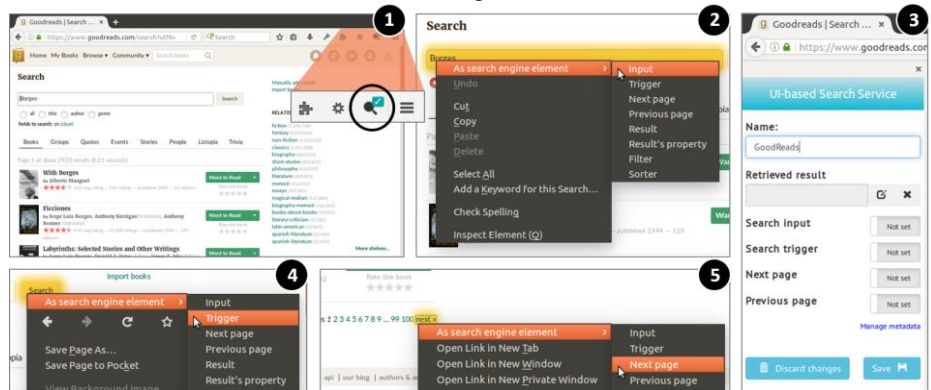

**Fig. 6**. Defining the input, trigger and pagination elements for a Search Service

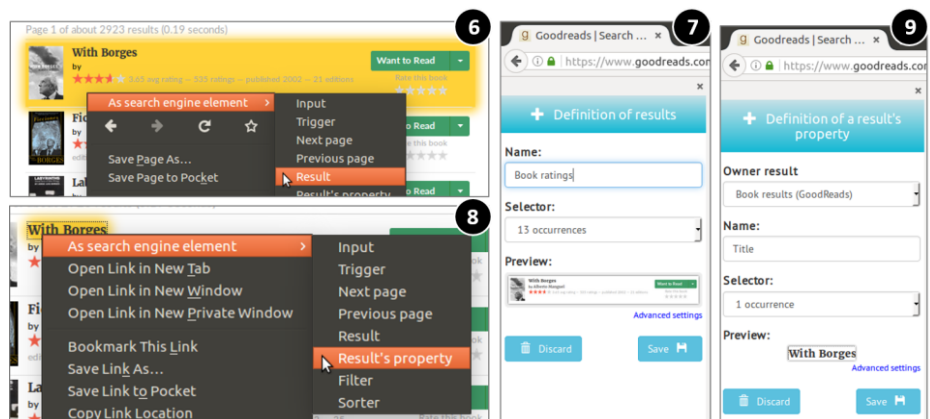

**Fig. 7**. Defining the expected results for a UI-based Search Service

Then, she specifies the kind of results the Search Service will retrieve, as shown in Figure 7. First, she selects an element in the DOM which represents the main container of the element he is expecting to have as a result. Such element is the highlighted one in step 6. When she chooses to define it as a "result" though the context menu options, a form is loaded in the sidebar (step7), where she must complete some required data. In this case, she names this kind of results as *Book rating* and selects one of the available selectors generated in relation to the available XPaths for the selected DOM element. This will allow her to choose more than one instance of *Book rating* in the same context (as shown in steps 6 and 8, there are other instances in the background). In a similar way, she should define the properties of the results that are of her interest, so these will be displayed when an ancillary search is performed. For instance, in steps 8 and 9 Amaru is defining the *Title* property for a *Book rating* result of the *GoodReads* Search Service. She repeats the last two steps for defining also a *Rating* property.

After the mandatory elements of a *UI-based Search Service* have been defined (input, more-results trigger and the expected structure of the results), the Search Services becomes available in the browser's context menu whether the user has highlighted any text in any Web page. In step 1 of Figure 8, Amaru has highlighted some text of a Wikipedia article and she is performing an ancillary search using it as keywords.

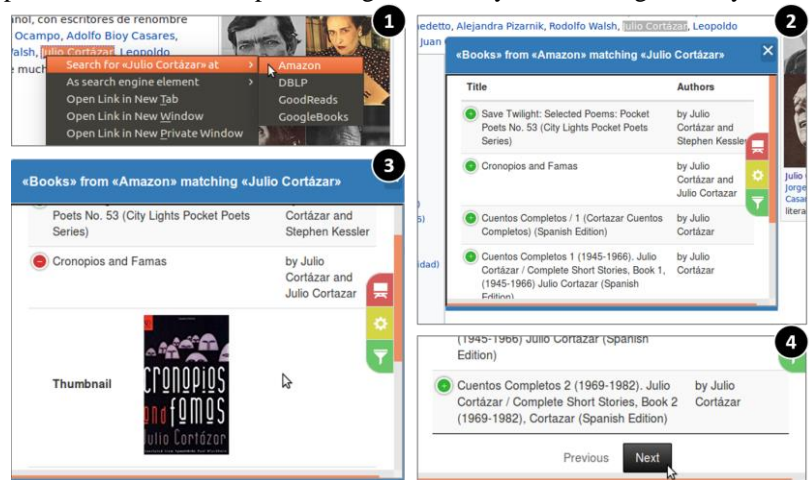

**Fig. 8:** visualizing the results of two ancillary searches

At that point, she can also use other previously defined services, for instance, *Amazon* or *Google Books*. When she clicks one of the menu items, let us say *Amazon*, a draggable panel appears in the middle of the screen, presenting the results of the search for the highlighted keywords. Now, as shown in step 2, she can access related books to *Julio Cortázar* and see their *Title* and *Authors*, but she can also access the remaining properties (in this case, the *Thumbnail*) by clicking the «+» button at the left of each row, which will display a section with the data hidden due to the lack of space. There are three fixed buttons at the right of the panel that allow her to: 1) change the kind of visualization – she is using the default one –; 2) to configure some parameters of the selected visualization, as the order or the priority of the columns in the responsive layout; and 3) to apply filters if any was defined. At the bottom of the visualization, as shown in step 4, there are navigation buttons so she can get more results.

Note that the use of the Search Service is not exclusive for the Wikipedia Web site; it is always available in the context-menu of the browser, no matter the Web site the user is navigating. Multiple ancillary searches with the same or diverse Search Services can be performed in a same context and using different keywords, as shown in step 6 and 7 of Figure 9. This way, Amaru can search for a second time, by selecting *Rayuela*, one of the books of *Julio Cortázar* listed in the first ancillary search' results popup. This time she is using the Search Service she defined for *GoodReads*, and she is accessing information that was not present in the results of the *Amazon* Search API.

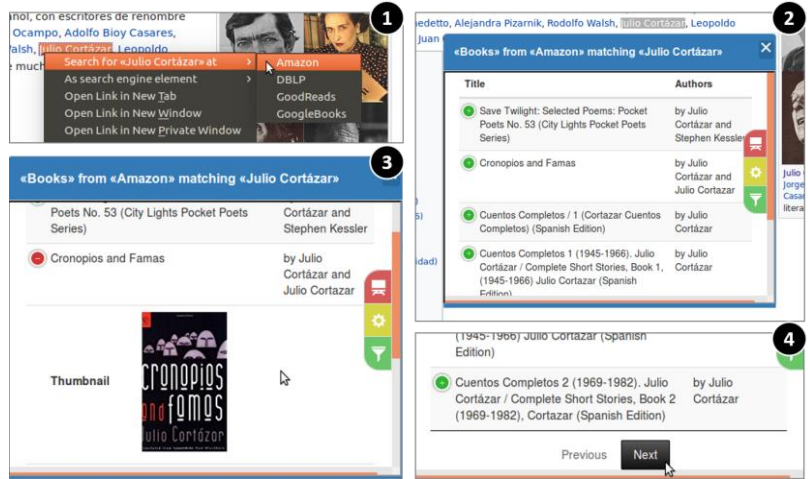

**Fig. 8:** visualizing the results of two ancillary searches

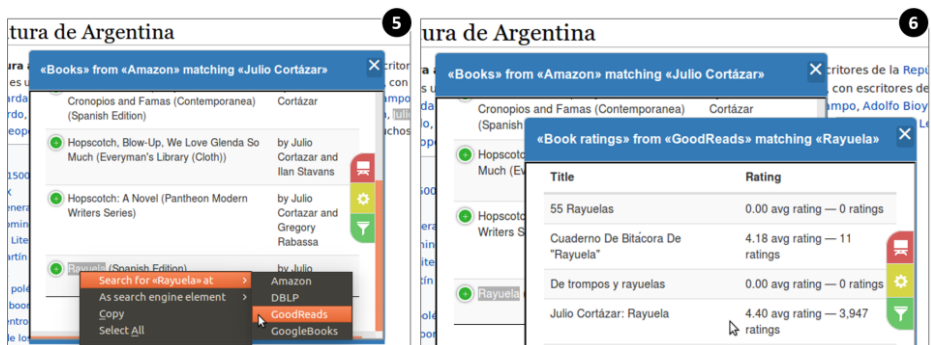

**Fig. 9:** visualizing the results of two ancillary searches

## 5 Evaluation

In this section, we present two evaluations of our approach. Section 5.1 proposes a validation by software construction, whose main objectives are to stress our Search Service model to know how it covers the existing search functionality provided by Web applications and also to measure the time consumed in real uses of these services. Section 5.2 presents a quantitative assessment to provide some understanding in relation to how our approach impacts on the user interaction.

### 5.1 Validation by construction

The instantiation of Search Services brought different challenges, especially considering that they are built by different UI components and consequently require different kind of interactions for automatically executing their associated behaviours. For instance, consider the 20 first sites in the top 500 by Alexa meeting the following conditions: 1) the interface is not Chinese or Russian by default; 2) take just one instance of the sites with same domain-name but different top-level domain; 3) only consider one instance using the same search engine (e.g. Msn redirects to Bing); 4) do not consider the one with no search engine (t.co). The 20 sites are Google, Youtube, Facebook, Yahoo, Wikipedia, Amazon, Live, Vk, Twitter, Instagram, Reddit, Linkedin, Ebay, Bing, Tumblr, Netflix, Wordpress, Microsoft, Aliexpress and Blogspot. In such a list, all the sites have a search engine but they are executable by different means.

By analysing the 20 sites, we can see that the involved UI controls differ in kind and quantity. There is no variation on the kind of DOM element used for entering the search's keywords, it is an input, but it is different for the remaining controls involved in the search process. Moreover, there is a site hiding the input until the user clicks on a concrete element of the DOM (Live.com), and the search execution strategies (see Section 3.3.1) are not always the same: 11 sites uses the *WriteAndClickToReload*; 5 the *WriteAndClickForAjaxCall*; 2 the *WriteForAjaxCall*. We successfully defined the service for those 18 of the list of 20 sites with such strategies; but the remaining 2 (Instagram and Live) required different ones that we are currently working on.

Back to the UI components, 17 of the full list of sites have a trigger element but they differ in kind of control: they were 10 buttons, 6 inputs and 1 anchor. Concerning pagination, 14 sites have a control for retrieving the following elements, but just 12 of them have a control for the previous ones. This is due to the way they handle results in the presentation layer; Instagram and Wordpress have a single DOM element that attaches more results in the results area, expanding its height. Facebook, Live, Vk, Twitter and Tumblr automatically retrieve more results when the user scrolls down to the bottom of the page. The remaining site, Microsoft, have a DOM element with this purpose but clicking on it redirects to a specialized form for searching a concrete kind of results, where the results of the first page were included. But such specialized searcher does not allow to change the search keywords; if you do that, you are redirected to the first results' page. Of the 14 sites with clickable paging elements, 8 of them cause the page to be reloaded while another 6 apply the changes via ajax-call. In this matter, sorting elements are present in 8 sites; 5 of them reload the document and 3 of them use ajax. Filter elements are defined for 16 sites, of which 11 reload and 5 use ajax.

The domain-specific abstraction of results is another issue to face. Most of the list of sites can retrieve more than one kind of result: news, images, videos, channels, playlists, people, pages, places, groups, applications, events, emails, etc. This is not just a problem for naming the results' kind, but for choosing a selector (an XPath) to retrieve the concrete type of element in the DOM (or all of them), whenever possible.

Regarding the performance of the searches, we logged the times for the 18 search engines our extension worked in a 15-inch notebook, with a resolution of 1366x768 pixels. The purpose of this analysis was to demonstrate that our tool allows the instantiation of the Search Services. We successfully defined a Search Service for each of the search engines, and we report below the times it took for executing an ancillary search, all of them: 1) from the same Web context[4], 2) searching for the same keywords (*Borges*) and 3) expecting to have results with just two textual properties: name and description (common to all the results). There were only two cases in which *Borges* did not produce any results, so the word was replaced by one of the same length: *Ubuntu*. For each test, we cleared the cached Web content, the offline content and user data. We also reloaded the target Web page to augment, for making it sure that no class was already instantiated and giving all the Search Services the same conditions before executing the search. The full search process took 7 seconds with a standard deviation of 5 seconds. The differences really depended on the strategy and the response time of the search engines. For instance, there were 3 cases in which it took between 14 and 18 seconds, while there were other 3 sites in which it took between 1 and 2 seconds. For a full list of logged times and a video, please visit the project's Web site[5].

### 5.2 Quantitative Assessment based on GOMS-KLM

We present in this section a quantitative assessment of search tasks based on the GOMS-Keystroke (KLM) model [10]. This formal model is used to evaluate the efficiency of interaction with a given software for a specific, very detailed, scenario. The resulting time is calculated by the summation of each user action, whose time are already known in this model. For example, the average time to perform the action *reach for mouse is of 0,40 sec, click on button is of 0.20 sec,* etc. Thus, by providing a detailed scenario of user actions with the Web browser and Web applications, it is possible to use GOMS-KLM to predict performance.

We focus this quantitative assessment for the case of ancillary searches. Using GOMS-KLM we have specified the scenario from Figure 1.b (the traditional way of performing ancillary searches) in order to compare it with our approach, which was introduced in Figure 2. We have split the whole scenario in six tasks: *i)* Visit ICWE2016's Web site, *ii)* select an author, *iii)* search for that author in DBLP, *iv)* select a resulting paper, *v)* search for that paper in Google Scholar, *vi)* point the mouse over the article's title. The results are shown in Table 1. Note that tasks *i, ii, iv* and *vi* are equivalents in both scenarios. However, the time required for tasks *iii* and *v*, the ones actually requiring to use the search engine provided by other Web sites, are quite faster using our approach. The whole scenario (involving two ancillary searches) takes 46,6 seconds without our approach. In opposition, using Search Services it takes just 18 seconds. It is interesting to note that performing a search using our approach and visualizing the results from any Web site would take only 1,5 secs (this time would be always the same, regardless both the current Web site and the target Search Service).

**Table 1. GOMS-KLM Results for both scenarios**

| Task | Time (without using SS) | Time (using SS) |
|---|---|---|
| 1. Go ICWE Web site | 8,7 secs | 8,7 secs |
| 2. Select target author | 2,6 secs | 2,6 secs |

---

[4] https://en.wikipedia.org/wiki/Argentine_literature
[5] https://sites.google.com/site/webancillarysearches/testing-in-top-sites

| 3. Perform first ancillary Search (DBLP) | 15,9 secs | 1,5 secs |
|---|---|---|
| 4. Select article | 2,6 secs | 2,6 secs |
| 5. Perform second ancillary Search (Google Scholar) | 15,7 secs | 1,5 secs |
| 6. Point target paper's title (Google Scholar) | 1,1 secs | 1,1 secs |
| **Total** | **46,6 secs** | **18 secs** |
| *Define both Search Services* | - | *39,2 secs* |

However, the reader may appreciate that for using these Search Services (for DBLP and Google Scholar) it was necessary to define them. With this in mind, we have defined also the GOMS-KLM scenarios related to the creation of both DBLP and Google Scholar Search Services. They were defined considering: *i)* input search form, *ii)* trigger search UI component, *iii)* abstract the search result into the concept "*Paper*" with the property *title*, iv) give a name to the Search Service and save it. All these tasks (whose required interaction was almost equal for both applications) take 19,6 seconds (for a basic Search Services, without filtering nor ordering options; features which were not required for the mentioned examples). This means that during the first time the user follows the scenario from Figure 2, it requires an extra time near to 40 seconds. However, this extra work is required only once, or even reduced to install an existing Search Services specification.

As a final comment, it is interesting to note that in the context of ancillary searches, the user would go back to the primary context, in this case, the ICWE's Web site. With our approach it would not require further interaction because the user already is in ICWE's Web site, meanwhile in a traditional scenario it will require further interaction.

## 6   Conclusions and Future works

In this paper, we presented an end-user driven approach for the customization of Web search tasks. The main objective is to get better support of ancillary searches, although the approach also reaches primary searches. Several contributions are made in this context. First, we propose an end-user support tool for the creation of Search Services based on the automatic execution of UI interaction required to perform searches in existing and third-party search engines. Second, we propose to transform search results into domain objects. Both together achieve a third contribution, which is the creation of new ways of interaction and visualization of results in-situ.

We fully supported our approach with already working tools, used in 18 existing Web applications as a way of validating our aims by software construction. In addition, we have shown that our approach is very convenient in terms of performance when users require complementary information for accomplishing their tasks.

Our approach still lacks a full end-user evaluation to measure the potential of adoption and the usefulness of in-situ visualizations of results. Another evaluation with end-users is necessary to demonstrate that the specification of Search Service is actually doable without programming skills. However, based on experts' judge, we strongly believe that, beyond some usability issues, the abstraction of Search Engine components should not be a limitation because of the seamless observed in existing search engines and their common use nowadays.

Beyond further evaluations and the improvement of our tools (e.g. to support more strategies in order to be compatible with other search engines), some other works are planned. First, although we foresee the usefulness of defining metadata for Search Services, we have not exploited it yet. For instance, this kind of information could be used for automatically perform searches in parallel given a particular context of information.

Collaboration is another aspect to be addressed. So far, we allow users to share service specifications by sending their corresponding files, but it is something we will investigate for reaching better results. Finally, domain-specific visualizers could improve how end-users interact with the information obtained by our search services.